\documentclass[a4paper]{jpconf}
\usepackage{graphicx}
\begin{document}
\title{Constituent quark models and pentaquark baryons}

\author{Kim Maltman}

\address{Dept. Math and Stats, York Univ.,
4700 Keele St., Toronto, ON CANADA M3J 1P3, and\\
CSSM, Univ. Adelaide, Adelaide, SA 5005 AUSTRALIA}


\begin{abstract}
We discuss certain general features of the pentaquark picture for the $\theta$,
its $\overline{10}_F$ partner, $\Xi_{3/2}$,
and possible heavy quark analogues. Models employing 
spin-dependent interactions based on either effective Goldstone
boson exchange or effective color magnetic exchange are also
used to shed light on possible corrections to the Jaffe-Wilczek
and Karliner-Lipkin scenarios. Some model-dependent
features of the pentaquark picture (splitting
patterns and relative decay couplings) are also discussed in
the context of these models.
\end{abstract}

\section{Introduction}
As a basis for the discussion which follows, I adopt the
following version of the current ``experimental situation''
with regard to reported pentaquark resonances:
\begin{itemize}
\item A significant number of medium-energy experiments
report positive evidence for the $\theta$. Its existence is thus plausible.
If these experiments are correct, the $\theta$ mass and width
are $m\sim 1540$ MeV and $\Gamma\sim O(1)$ MeV. See Ref.~\cite{hicksexprev} for
a critical assessment of both positive and negative
experimental searches, as well as references to the
experimental literature.
\item The NA49 exotic $I=3/2$, $\Xi$ ($\Xi_{3/2}$) signal 
with $m\sim 1860$ MeV and narrow width~\cite{na49} has not
been confirmed~\cite{na49neg}, and is hard to reconcile with 
high statistics $\Xi$ production experiments~\cite{na49criticism}. 
Though a $\Xi_{3/2}$ partner is unavoidable if the $\theta$ lies in a 
$\overline{10}_F$, in view of the experimental situation,
no assumption will be made about its mass.
\item The H1 report of an anticharmed baryon at $3099$ MeV~\cite{H1} 
is also unconfirmed~\cite{ZEUSH1}, so no assumptions will
be made about the masses of any possible heavy pentaquark states.
\end{itemize}

A low-lying, relatively narrow ($O(10 \ {\rm MeV})$), $\overline{10}_F$,
$J^P=1/2^+$, $S=+1$ exotic baryon is natural in the chiral soliton model 
(CSM)~\cite{dpp,ekp}. In the quark model,
the ground state of the exotic system
would have negative parity ($P=-$) if all 
five particles were in a symmetric relative s-wave configuration.
Whether such a configuration represents the true model
ground state, however, is a dynamical question.
Exciting one of the four light ($\ell =u,d$) quarks into a 
p-orbit, reducing the $\ell^4$ spatial 
symmetry from $[4]_L$ to $[31]_L$, allows spin-flavor-color configurations 
with significantly increased hyperfine attraction which
were previously Pauli-forbidden. If the
increased attraction is greater than 
the p-wave excitation cost, the ground state will have $P=+$.
A dynamical calculation is needed to determine
which of the two effects wins out. 

Two proposals of possible $P=+$ quark model configurations for
the $\theta$ have received considerable attention, and 
are also useful for framing the discussions below. 
The Jaffe-Wilczek (JW) ansatz for the $\theta$~\cite{jw} involves
two $I=J=0$, $C=\bar{3}$ diquark pairs, coupled
to color $C=3$, with only confinement forces between the 
diquarks and $\bar{s}$.
Bose statistics for the diquarks imply a relative p-wave between
them, hence $P_\theta =+$. 
The Karliner-Lipkin (KL) scenario for the $\theta$~\cite{kl} involves
one JW diquark and one $ud\bar{s}$ triquark. In the triquark,
the spin and color of a JW diquark have been flipped and anti-aligned
with those of the $\bar{s}$, producing triquark 
quantum numbers $I=0,J=1/2,C=3$. The ansatz is motivated by the 
color-magnetic-exchange model, in which the 
hyperfine attraction is stronger for the KL than for the JW correlation.
A relative diquark-triquark p-wave will yield $P_\theta =+$.
In both scenarios, the relative intercluster p-wave 
is assumed to make cross-cluster interaction and antisymmetrization
(AS'n) effects negligible.

The ``GB model'' is a model with spin-dependent 
interactions generated by effective Goldstone boson exchange. 
The spin-dependent part of the Hamiltonian is
\begin{equation}
H_{GB}\sim -\, \sum_{i<j}\vec{\sigma}_i\cdot\vec{\sigma}_j
\ \lambda^a_{F,i}\lambda^a_{F,j}
{\frac{f^a(\vec{r}_{ij})}{m_im_j}}\label{hgb}\end{equation}
with $\vec{\sigma}_i$ the Pauli spin matrices, $m_i$ the constituent
quark mass, $\lambda^a_{F,i}$ the $SU(3)_F$ Gell-Mann matrices,
and $f^a(\vec{r})$ as given in Ref.~\cite{gbmodel}. 
The JW diquark is by far the 
most attractive $qq$ correlation for $H_{GB}$.
If the sum in Eq.~(\ref{hgb}) runs only over quark labels~\cite{gbmodel}, 
the model also matches
the JW assumption for the $\bar{s}\ell$ interactions.
Assuming an orbital excitation cost
similar to that in the ordinary baryon sector of the model,
the gain in hyperfine energy in the $[31]_L$ configuration 
leads to a $P=+$ ground state, one which should 
be dominated by the JW correlation. The model thus
provides a dynamical setting for studying (i) corrections
to the strict JW scenario (e.g., cross-cluster
AS'n/interaction effects, configuration mixing)
and (ii) the pattern of excitations above the JW-correlation-dominated
ground state.

The ``CM model'' is a model with spin-dependent interactions generated
by effective color magnetic exchange. The spin-dependent part
of the Hamiltonian is
\begin{equation}
H_{CM}\sim -\sum_{i<j=1,\cdots ,5}\vec{\sigma}_i\cdot\vec{\sigma}_j\ 
F^a_{i} F^a_{j}
{\frac{f_{CM}(\vec{r}_{ij})}{m_im_j}}\label{hcm}\end{equation}
with $\vec{\sigma}_i$ and $m_i$ as above,
$F^a_i=\lambda^a_i/2$ ($-{\lambda^{a}}^*_i/2$) for 
quarks (antiquarks) the $SU(3)_c$
Gell-Mann matrices and $f_{CM}(\vec{r}_{ij})$
a smeared delta function.
The JW diquark is again the most attractive $qq$ correlation, 
but, as noted above, the more complicated KL configuration 
lies lower than the JW configuration.
The CM model thus provides
a dynamical setting for studying (i) corrections to the KL
scenario and (ii) the expected excitation pattern
above the CM ground state. An important
correction results from the fact 
that the same $\bar{s}\ell$ interactions
responsible for lowering the KL triquark hyperfine energy
also mix the JW and KL corelations, producing a configuration
significantly lower than either the JW or KL correlations
separately. This configuration turns out to be
a roughly equal admixture of the two correlations~\cite{jm}.

\section{Results}
Space constraints prevent discussion of all points covered
in the conference talk. (See the conference website and
Refs.~\cite{jm,kmnstar04,kmheavy} for more details; 
Ref.~\cite{kmheavy}, in particular, gives an explicit definition of the
dimensionless form used in quoting results below.)

{\it (a) The $\theta$ mass and pentaquark widths in the quark model:}
Models such as the GB and CM models, with parameters
fixed from the ordinary baryon sector, cannot 
predict exotic pentaquark masses
because of the absence of terms representing the response
of the non-perturbative vacuum to the
hadronic constituents. This response will differ 
depending on the number of constituents. This effect will
be absorbed by the additive constant
used to set the overall scale in each sector; the constant
should thus be different for ordinary and pentaquark baryons,
just as it is for mesons and baryons. 
In the bag model, where some modelling of
vacuum response {\it does} occur, the difference in
1-body energies between a single $q^6$ bag and two isolated $q^3$ bags
is an extremely sensitive function of the $B^{1/4}$~\cite{kmbag},
suggesting it is impossible to anticipate in advance
the size, or even sign, of this effect. Thus, the fact that
quark model calculations for $m_\theta$ {\it using parameter 
values fixed from the ordinary baryon spectrum} yield masses 
$100-300$ MeV or so too high does not rule out the validity of
a pentaquark picture for such states. Other
observables (widths, splittings, relative decay
couplings) must be used to test the models.

If the $\theta$ is well described as a pentaquark,
the initial and final states in $\theta \rightarrow NK$
have the same number of constituents. The decay 
thus proceeds by rearrangement, or ``fall-apart'', and should have a
width $\sim \hat{\Gamma}S$, where $\hat{\Gamma}$ is
the width for a $KN$ potential scattering resonance with
mass $m_\theta$ in an attractive potential of roughly
hadronic size, 
and $S$ is a spectroscopic, or squared-overlap, factor~\cite{jm}.
A single-scale s-wave attractive potential too weak to bind
does not produce resonant behavior~\cite{jw}; hence
a $J^P=1/2^-$ assignment for the $\theta$ is impossible.
For a p-wave (d-wave) decay, $\hat{\Gamma}\sim 200$ (respectively, 20) MeV.
The spin-flavor-color (JFC) part of the overlap suppression is 
1/24 for the JW correlation and $\sim 1/25$ 
for the optimized JW/KL combination in the CM model~\cite{jm}. 
With additional suppression from the spatial overlap, $\theta$ widths $\sim$ 
a few to several MeV are expected for $P=+$ ground states
involving such correlations. As for the CSM, some fine tuning is needed
to reduce this to $\sim 1$ MeV or less.
(A very narrow $\theta$ is natural for a $J^P=3/2^-$
assignment, where a d-wave $NK$ decay is required. However,
at least in the GB and CM models, such a ground state assignment
is impossible~\cite{jm}.)

The JFC ``overlap'' suppression effect
is rather general in the quark model approach.
In the soft-$K$/PCAC approximation, the
$\theta\rightarrow nK^+$ coupling is 
$\propto\langle n\vert \bar{s}i\gamma_5 u\vert \theta\rangle$.
Three of the light quarks in the $\theta$ must project onto
the $n$ and the remaining constituent $\ell$
and $\bar{s}$ be annihilated by the pseudoscalar
density with $K^+$ quantum numbers. The JFC 
factors are the same as would occur in a $\theta$ to $NK$ overlap 
calculation using a constituent $\bar{s}u$ picture for the
$K^+$. A model calculation~\cite{oka04}
confirms the reduction of $\Gamma_\theta$ to $\sim 10$ MeV 
for the JW correlation, before inclusion of spatial overlap effects,
and can be interpreted as a model-dependent version of the
soft-$K$/PCAC estimate.
As stressed in Ref.~\cite{closezhao}, for $P=+$ pentaquarks
with p-wave fall-apart decays, decay coupling ratios
should be given by the corresponding overlap ratios squared.

{\it (b) Corrections to the JW and KL scenarios in the GB and CM models:}
For the GB model, $\langle H_{GB}\rangle$ in the strict JW limit
is $-16$. The lowest-lying $P=+$ state 
has JW quantum numbers (which, up to a $J^P=3/2^+$ spin-orbit
partner, are also the CSM ground state quantum numbers). 
Cross-cluster interaction/AS'n and configuration
mixing lower the actual ground state expectation
to $-21.9\pm 1.3$~\cite{jm}. The JW approximation
is reasonable, but not exact. The ground state expectation {\it is}, 
however, the same in the $\theta$ and heavy $\bar{Q}\ell^4$
($\bar{Q}=\bar{c},\bar{b}$) sectors, as in the JW scenario. 
With the JW estimate for the 1-body contribution to 
$m_{\theta_{c,b}}-m_\theta$~\cite{jw},
the JW prediction of strong interaction stable $\theta_{c,b}$
is thus preserved.

For the CM model, in the $\theta$ sector, the optimized combination
of JW and KL correlations provides a good approximation to the
lowest $P=+$ eigenvalue~\cite{jm}, which again occurs for 
the channel with CSM quantum numbers. In the heavy 
sector, the $\bar{Q}\ell$ interactions are strongly suppressed by the 
$1/m_{\bar{Q}}$ factor in $H_{CM}$. This (i) decreases the 
KL triquark hyperfine attraction, making the
KL diquark-triquark configuration higher in energy than
the JW correlation for both $Q=c,b$, and (ii) strongly suppresses 
KL-JW mixing.
The heavy pentaquark $P=+$ ground state should thus be
dominated by the JW correlation. The 
JW expectation is $-4$, c.f. the actual ground
state expectation $-3.48\pm 0.04$~\cite{kmheavy}. 
The JW approximation 
is again reasonable, but not exact. 
The KL scenario for the heavy quark analogues~\cite{klheavy}
assumes the same triquark-diquark clustering
as for the $\theta$ persists in the heavy system.
This assumption is {\it not} compatible with the CM model
effective interactions, which motivated the original triquark-diquark
proposal. The KL heavy sector ansatz thus leads to an overestimate
of the $\theta_{c,b}$ masses in the CM model. Modified estimates
put the $\theta_{c,b}$ close to the $ND$, $NB$ strong decay
thresholds~\cite{kmheavy}.

{\it (c) Splittings and relative decay couplings in the light
and heavy $P=+$ $\bar{Q}\ell^4$ sectors:}
Though one cannot predict absolute pentaquark masses 
in the quark model approach, splittings are independent of
any additive constant and hence {\it are} predictable.
Predictions for
a given $\bar{Q}\ell^4$ sector (with $Q=s,c,b$ fixed)
are particularly immune to uncertainties in treating
1-body energies, and are on the same footing as analogous
predictions for the ordinary baryon sector. The relative
ordering of states is fixed by the JFC structure of the
effective operators, with (in a given sector) an overall
scale, dependent on the size of the pentaquark states.
{\it Ratios} of overlap factors relevant to the decays to $BM$,
with $B$ the relevant ground state baryon and $M$ the relevant
vector or pseudoscalar meson, are also well-determined~\cite{kmheavy}.

For the GB model, the first excitation above the $P=+$,
$\bar{Q}\ell^4$ ground state, for all of $Q=s,c,b$,
is a degenerate pair with $(I,J_q)= (1,1/2),\ (1,3/2)$,
where $J_q$ is the total intrinsic spin. The 
excitation energy is $\sim 130-170$ MeV. No other states
are expected within $\sim m_\Delta -m_N$ of
the ground state. For the $(1,1/2)$ state, the combination
of phase space and squared-overlaps enhances the width
to $NK$ by a factor of at most $\sim 12$ c.f. the $\theta$,
making this prediction potentially problematic. A similar,
potentially problematic, state with $(F,I,J^P)=(27,1,3/2^+)$,
mass $\sim 60$ MeV above the $\theta$, and relatively narrow 
($37-66$ MeV) $NK$ width, is also
predicted in the CSM~\cite{ekp}.
For the CM model, in the $\ell^4\bar{s}$ sector, a single
low-lying $(I,J_q)=(1,1/2)$ excitation is predicted $\sim 80$ MeV
above the $\theta$. The next excitation lies $\sim 160$ MeV higher.
Phase space and the overlap factor for $(1,1/2)$ state's decay 
to $NK$ are again such that a relatively narrow width is expected.
Even more problematic are the predictions in the heavy sector, where {\it five}
low-lying excitations are expected, with
$(I,J_q)=(0,1/2)$, $(1,1/2)$, $(0,3/2)$, $(1,3/2)$, $(1,1/2)$,
the first $\sim 80-90$ MeV above the
ground state, and the last $\sim 130-150$ MeV above the ground
state~\cite{kmheavy}. Such a dense, low-lying excitation spectrum seems
unlikely, but would be a striking prediction of the CM model,
if confirmed. None of the overlap ratios are such as to allow 
any of these low-lying states to be experimentally undetectable
in 2-body decay modes, provided at least one of either them, or 
the ground state, is experimentally accessible~\cite{kmheavy}.

For the GB model, the JW dominance of the ground state and
decrease in JW diquark hyperfine energy under $d\leftrightarrow s$
imply a reduction in $\Xi_{3/2}$ hyperfine energy c.f.
the $\theta$. An estimate of the splitting, using only the $[4]_{FJ}$
component in the $\Xi_{3/2}$ wavefunction, 
yields $m_{\Xi_{3/2}}=1962$ MeV~\cite{stancu04}.
Allowing other wavefunction components (known to be
small for the $\theta$, but expected to be more important for the $\Xi_{3/2}$)
will lower this value. In the KL scenario, the
$\theta$ and $\Xi_{3/2}$ have almost identical
$\langle H_{CM}\rangle$ values, leading to 
$m_{\Xi_{3/2}}\simeq 1720$ MeV~\cite{kl}. 
The lowest hyperfine eigenvalues in the 
$\theta$ and $\Xi_{3/2}$ channels turn out to share this near equality,
even though the ground state 
is not well approximated by the KL correlation. The CM model thus
produces a significantly smaller $\Xi_{3/2}-\theta$ splitting
than does the GB model.

\end{document}